\documentclass[aps,prl,showpacs,twocolumn,floatfix,a4paper,superscriptaddress]{revtex4}

\usepackage{amsmath,amsfonts,amssymb,graphics,graphicx,epsfig,color,times,bbm,psfrag}

\DeclareMathOperator{\spec}{spec}

\begin{document}

\bibliographystyle{apsrev}

\newcommand{\R}{\mathbbm{R}}
\newcommand{\rr}{\mathbbm{R}}
\newcommand{\nn}{\mathbbm{N}}
\newcommand{\cc}{\mathbbm{C}}
\newcommand{\ii}{\mathbbm{1}}
\newcommand{\id}{\mathbbm{1}}

\newcommand{\tr}{{\rm tr}}
\newcommand{\gr}[1]{\boldsymbol{#1}}
\newcommand{\be}{\begin{equation}}
\newcommand{\ee}{\end{equation}}
\newcommand{\bea}{\begin{eqnarray}}
\newcommand{\eea}{\end{eqnarray}}
\newcommand{\ket}[1]{|#1\rangle}
\newcommand{\bra}[1]{\langle#1|}
\newcommand{\avr}[1]{\langle#1\rangle}
\newcommand{\G}{{\cal G}}
\newcommand{\eq}[1]{Eq.~(\ref{#1})}
\newcommand{\ineq}[1]{Ineq.~(\ref{#1})}

\newtheorem{theorem}{Theorem}
\newtheorem{proposition}{Proposition}
\newtheorem{lemma}{Proposition}
\newtheorem{definition}{Definition}
\newtheorem{corollary}{Corollary}

\title{The optimal cloning of quantum coherent states is non-Gaussian}

\author{N.\,J.~Cerf}
\email{ncerf@ulb.ac.be}
\affiliation{QUIC, Ecole Polytechnique, CP 165, Universit\'e Libre de Bruxelles, 1050 Brussels, Belgium}
\author{O.~Kr\"uger}
\email{o.krueger@tu-bs.de}
\affiliation{Institut f\"ur Mathematische Physik, Technische Universit\"at %
    Braunschweig, Mendelssohnstra{\ss}e 3, 38106 Braunschweig, Germany}
\author{P.~Navez}
\affiliation{QUIC, Ecole Polytechnique, CP 165, Universit\'e Libre de Bruxelles, 1050 Brussels, Belgium}
\author{R.\,F.~Werner}
\affiliation{Institut f\"ur Mathematische Physik, Technische Universit\"at %
    Braunschweig, Mendelssohnstra{\ss}e 3, 38106 Braunschweig, Germany}
\author{M.\,M.~Wolf}
\email{michael.wolf@mpq.mpg.de}
\affiliation{Max-Planck-Institut f\"ur Quantenoptik, %
    Hans-Kopfermann-Stra{\ss}e 1, 85748 Garching, Germany}

\begin{abstract}
We consider the optimal cloning of quantum coherent states with single-clone
and joint fidelity as figures of merit. Both optimal fidelities are attained
for phase space translation covariant cloners. Remarkably, the joint fidelity 
is maximized by a Gaussian cloner, whereas the single-clone fidelity can be
enhanced by non-Gaussian operations: a symmetric non-Gaussian 1-to-2 cloner can
achieve a single-clone fidelity of approximately 0.6826, perceivably higher than
the optimal fidelity of 2/3 in a Gaussian setting. This optimal cloner can be
realized by means of an optical parametric amplifier supplemented with a
particular source of non-Gaussian bimodal states.
Finally, we show that the single-clone fidelity of the optimal 1-to-$\infty$
cloner, corresponding to a measure-and-prepare scheme, cannot exceed 1/2. This
value is achieved by a Gaussian scheme and cannot be surpassed even with
supplemental bound entangled states.
\end{abstract}


\maketitle


\newcommand\fjoint{f_\text{joint}}

The no-cloning theorem states that there is no quantum apparatus capable of
perfectly duplicating an arbitrary input state \cite{nocloning}. This is a
direct consequence of the linearity of quantum mechanics and a fundamental
difference between classical and quantum information. This theorem enables one
of the most promising applications of quantum information theory, namely secure
quantum key distribution. Moreover, the impossibility of perfect cloning
machines is intimately connected to other impossible tasks in quantum mechanics
\cite{RFWimpossiblemachines}.

Soon after the observation of the no-cloning theorem as a
fundamental feature of quantum mechanics the question arose how
well an approximative cloning machine could work. For the case of
universal cloning of finite dimensional pure states this question
was addressed and answered in \cite{HB,GM,BDEFMS,W,KW,BH98,C98}. There,
the figure of merit was the fidelity, i.\,e.\ the overlap between
hypothetically perfect clones and the actual output of the imperfect cloner.
In particular, it was shown that judging single clones leads to the same
optimal cloner as when comparing the joint output with a tensor product
of perfect clones \cite{W,KW}.

Recently, more and more attention has been devoted to continuous variable
systems, especially to states with Gaussian Wigner function -- so called
Gaussian states. Besides their outstanding importance in quantum optics and
quantum communication, in particular quantum cryptography \cite{G03},
they provide a closed testbed within which many of the otherwise hardly tractable problems in
quantum information  become feasible. Restricting to the Gaussian world,
i.\,e.\ to Gaussian operations on Gaussian states, led for instance to
solutions to otherwise open problems in the theory of entanglement measures
\cite{GEOF}, quantum channels \cite{SEW} and secret key distillation
\cite{NBCLSA}. Similarly, the problem of cloning in particular coherent
states by Gaussian operations has been addressed in \cite{L,CIR}.
The obtained cloner was shown to be optimal within the class of Gaussian
operations by exploiting the connection with state estimation \cite{CI}.
However, it remained unclear whether Gaussian operations really lead to the
optimum, even under the assumptions typically made in the literature such as
phase space translation covariance or output symmetry.

The present Letter is concerned with the problem of optimally cloning coherent
states without imposing any restrictions on the cloning operation. After
recalling some preliminaries, we will prove that w.\,l.\,o.\,g.\ one can
restrict to covariant cloners, for which a powerful characterization will be
provided.  Based on this, we will show that, in contrast to the finite
dimensional case, the optimal cloner depends on whether we judge single clones
or test the clones jointly. Surprisingly, in the latter case the known Gaussian
cloners turn out to be optimal, whereas with respect to the single-clone
fidelity, non-Gaussian operations can perform better.

The problem of finding the optimal cloner reduces to finding the dominant
eigenstate of an appropriate operator. For the optimal 1-to-2 cloner this
eigenstate is directly linked to an optical implementation: it is the bimodal
state of light that has to be injected on the idler mode of an optical
parametric amplifier and the input port of a beam splitter. We envision that a
few-photon approximation of this cloner, only sub-optimal but yet non-Gaussian,
might be feasible, making it possible to experimentally demonstrate this
fidelity enhancement.

In addition, we will show that a 1-to-$\infty$ cloner based on a
measure-and-prepare scheme cannot exceed a fidelity of 1/2, not
even with supplemental bound entangled states.
Extended discussions of the mathematical details \cite{KWW} and the quantum
optical aspects \cite{NC} will be reported elsewhere.

\paragraph{Phase space and coherent states.}

Consider a system of $n$ harmonic oscillators with respective canonical
operators, or optical field quadratures, $(Q_1,P_1,\ldots,Q_n,P_n)=:R$ and the
corresponding phase space $\Xi=\mathbb{R}^{2n}$, which is equipped with an
antilinear symplectic form $\sigma(\xi,\eta)$. Translations in this phase space
are governed by the \emph{Weyl} or \emph{displacement operators}
$W_{\xi}=e^{i\, \sigma(\xi,R)}$, $\xi\in\Xi$, which in turn obey the Weyl
relations
\be\label{eq:Weyl}
W_{\xi}\, W_{\eta}=e^{-\frac{i}{2} \sigma(\eta,\xi)}\,
W_{\xi+\eta}, \text{ where } \sigma = \bigoplus_{i=1}^n
\left(\begin{array}{cc}
  0 & 1 \\
  -1 & 0 \\
\end{array}\right)
\ee
implements the symplectic form via $\sigma(\xi,\eta) = \xi^T\cdot\sigma\cdot\eta$.
Tensor products in Hilbert space correspond to direct sums in phase space, and
in particular $\bigotimes_i W_{\xi_i^{}} = W_{\bigoplus_i\xi_i^{}}$ where each of the
$\xi_i\in\mathbb{R}^2$ belongs to a single mode.

The expectation values of the Weyl operators completely determine a state, and
the resulting function, which is the Fourier transform of the Wigner function,
is called the \emph{characteristic function}. For a \emph{coherent state}, it is
a Gaussian of the form \be \chi(\xi) = \tr[\rho\, W_{\xi}] =
e^{-\frac14\xi^T\cdot\gamma\cdot\xi-i\, d^T\cdot\xi}, \ee with covariance
matrix $\gamma=\ii$ and displacement vector $d$. Coherent states are
translations of the harmonic oscillator ground state $W_\xi
|0\rangle=|\xi\rangle$ with $d=\sigma\cdot\xi$. In quantum optical settings,
position and momentum coordinates correspond to the real and imaginary parts of
the complex field amplitude.

\paragraph{Figures of merit.}

The \emph{fidelity} quantifies how close two states $\rho_1$ and $\rho_2$ are
\cite{fidelity}. Here, we only consider the case of pure input states, so we
can simply set $f(\rho_1,\rho_2)=\tr[\rho_1\, \rho_2]$. A $1$-to-$n$ cloning
transformation $T$ (a ``cloner'' for short) by definition takes systems in the
pure input state $\rho$ into $n$ systems whose state is close to $n$
copies of $\rho$. We can express this by requiring the
fidelity
\begin{equation}\label{fjoint}
    \fjoint(T,\rho)=\tr\bigl[T(\rho)\, \rho^{\otimes n}\bigr]
\end{equation}
to be as large as possible. This is a very demanding criterion, as it also
evaluates whether the clones are nearly independent. Instead, we might just
evaluate the quality of an individual clone, say the $i^\text{th}$,
\begin{equation}\label{fi}
    f_i(T,\rho)
    =\tr\bigl[T(\rho)\, (\ii\otimes\cdots \otimes\ii\otimes\rho^{(i)}\otimes\ii
               \otimes\cdots\otimes\ii)\bigr],
\end{equation}
where the upper index denotes the position in the tensor product. Since a
single such fidelity can be trivially put to one by copying the input 
onto the $i^\text{th}$ clone, we have to maximize a weighted sum $\sum_i\lambda_if_i(T,\rho)$ with positive weights $\lambda_i$.

Further options arise from the choice of the set of states $\rho$ that 
we want to clone optimally. Here, we consider the family of coherent states $\rho=\ket\xi\bra\xi$, with $\ket\xi=W_\xi \ket0$. 
We define $\fjoint(T)$ and $f_i(T)$ 
as the respective \emph{worst-case
fidelities}, i.\,e.\ the minima of (\ref{fjoint}) and (\ref{fi}) 
over all coherent states $\rho$. Note that this is different from the usual
case of universal cloners in finite dimensional Hilbert spaces, 
where one considers the minimum with respect to all pure states. 
This is connected to the infinite number of
dimensions of the continuous variable Hilbert space: Even
minimizing (\ref{fjoint}) or (\ref{fi}) over all pure squeezed Gaussian states
(a larger though still very small subset of all states) would already 
yield a zero fidelity for all $T$.

Our goal is thus to find the optimal worst-case joint fidelity
\begin{equation}
\fjoint= \sup_T \fjoint(T)=\sup_T \inf_{\rho\in \text{coh}} \fjoint(T,\rho)
\end{equation}
as well as the convex set of achievable $n$-tuples of single-clone fidelities
$\bigl(f_1(T),f_2(T),\ldots,f_n(T)\bigr)$ 
as $T$ varies over all cloners.
This is simplified by the fact that both fidelities are invariant under
displacements in phase space, so we can choose the optimal cloner to be
covariant. Consequently, they are optimal with respect to both worst-case and
average fidelities.

\paragraph{Covariance.}

Let $T$ be a 1-to-$n$ cloning map. If displacing the input in
phase space is equivalent to displacing the outputs by the same
amount, then $T$ is called (displacement) covariant:
\be\label{eq:cov}
    T(\rho)=W_\xi^{\otimes n \dag}\, T\, \bigl(W_\xi^{}\, \rho\,
        W_\xi^\dag \bigr) W_\xi^{\otimes n}
        \equiv T_\xi^{}(\rho)
\ee
for all $\xi$ and $\rho$, where we have defined the shifted cloner $T_\xi$
for later reference. The cloners investigated in
\cite{L,CIR} were restricted to be covariant. However,
this need not be assumed, but rather comes out as a property of the
optimal cloners. As in the case of cloning of finite-dimensional
systems \cite{KW}, the core of the argument is averaging over the
symmetry group: we have, for $f=\fjoint$ or
$f=\sum_i\lambda_i f_i$, respectively,
\begin{equation}\label{av}
\begin{split}
    f(T) &= \inf_\xi f(T,\ket\xi\bra\xi)
        \leq {\bf M}_\xi f(T_\xi,\ket0\bra0) \\
    &= f({\bf M}_\xi T_\xi,\ket0\bra0)
        = f({\bf M}_\xi T_\xi).
\end{split}
\end{equation}
Here ${\bf M}_\xi$ stands for ``mean with respect to $\xi$'' and is implemented
by an invariant mean \cite{invmean}. So, the averaged and thus covariant cloner
is at least as good as $T$ for all $T$, and we can restrict the search to the
covariant case.

Note that the output of such cloners could be singular for this phase space
average. A detailed argumentation shows, however, that this is not optimal for
the fidelities considered here \cite{KWW}.

\paragraph{Optimizing covariant cloners.}

In the Heisenberg picture, (the adjoint of) a covariant cloner maps Weyl
operators onto multiples of Weyl operators,
\begin{equation}\label{chiT}
    T_\ast(W_{\xi_1,\ldots,\xi_n})
    =t(\xi_1,\ldots,\xi_n)\, W_{\sum_i\xi_i^{}},
\end{equation}
where $\xi_i$ is the pair of phase space variables of the $i^\text{th}$ clone.
In terms of characteristic functions of input and output states, $t$ acts as a
characteristic function of the cloner:
\begin{equation}\label{Tchi}
    \chi_\text{out}(\xi_1,\ldots,\xi_n)
        = t(\xi_1,\ldots,\xi_n)\,
		\chi_\text{in}\bigl(\textstyle\sum_i\xi_i^{}\bigr).
\end{equation}
The condition of complete positivity requires that $t$ is the characteristic
function of a state $\rho_T$, plus a fixed linear transformation \cite{DVV}. We
call a cloner \emph{Gaussian} if $t$ has a Gaussian form and it thus maps
Gaussian input states onto Gaussian output states.

Since fidelities are linear in $T$, and hence linear in $\rho_T$, they can
be expressed as expectation values of linear operators:
\begin{equation}\label{Fops}
    f(T,\rho)=\tr[\rho_T\, F].
\end{equation}
The appropriate operators $F_\text{joint}$ and $F_i$ do not depend on $T$,
which allows us to reduce the supremum of the left-hand side  of (\ref{Fops})
to finding the state $\rho_T$ (hence the map $T$) corresponding to the largest
eigenvalue of $F$. This is the core of our method.

Physically, the state $\rho_T$ is directly related (up to a suitable symplectic
transformation) to the bimodal state that needs to be injected on the idler
mode of an optical parametric amplifier together with the input port of a beam
splitter in order to realize the cloner $T$ (see below).

\paragraph{Optimal fidelities.}

Since by Eq.(\ref{av}) the maximum fidelities are reached by covariant cloners, we can restrict
the further discussion to a vacuum input state $\rho=\ket0\bra0$. 
For Gaussian input states, the operators $F$ in Eq.~(\ref{Fops}) are
themselves Gaussian, so that the respective fidelities $f$ 
are optimized by Gaussian pure states $\rho_T$, hence by Gaussian cloners $T$.
Consequently, the joint fidelity $f_\text{joint}(T) =
f_\text{joint}(T,\ket0\bra0) = \tr\bigl[\rho_T\, F_\text{joint}\bigr]$ is
maximized by a Gaussian cloner. The maximum fidelity is given by the largest
eigenvalue of an appropriately defined operator $F_\text{joint}$, that is
\begin{equation}
\sup_T\ \fjoint(T) = \max \spec (F_\text{joint}) = \frac{1}{n}. 
\end{equation}
Thus, the unique optimal cloner in this case
is the know Gaussian cloner of \cite{L,CIR,CI}.

\begin{figure}
\psfrag{f1`}[cl][cl]{\tiny $f_1$}
\psfrag{f2`}[Bc][Bc]{\tiny $f_2$}
\psfrag{1.}[tc][tc]{\tiny 1}
\psfrag{0.9998}[tc][tc]{\tiny 0.9998}
\psfrag{0.0002}[cr][cr]{\tiny 0.0002}
\psfrag{0.0004}[cr][cr]{\tiny 0.0004}
\psfrag{f1}{$f_1$}
\psfrag{f2}{$f_2$}
\psfrag{0}[Br][Bl]{0}
\psfrag{1}[tc][tc]{1}
\psfrag{a1}[tc][tc]{1/2}
\psfrag{b1}[tc][tc]{2/3}
\psfrag{a2}[Br][Br]{1/2}
\psfrag{b2}[Br][Br]{2/3}
\includegraphics[width=\columnwidth]{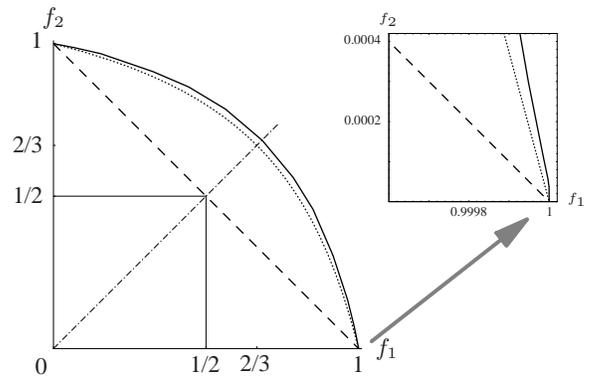}
\caption{Achievable pairs $(f_1,f_2)$ of single-clone fidelities in 1-to-2
cloning of coherent states. The dots represent the optimal Gaussian cloner,
while the solid curve indicates optimal non-Gaussian operations.  Fidelities in
the lower left quadrant are accessible to measure-and-prepare schemes.
Classical mixtures of the two ``trivial'' cloners fall onto the dashed line.
The dash-dotted diagonal marks symmetric cloners, with intersection points
corresponding to the classical, best Gaussian, and optimal cloning,
respectively. The inset shows the infinite slope at $f_1=1$ for non-Gaussian
cloners as opposed to the Gaussian case.}\label{pic}
\end{figure}

For the single-clone fidelity, we have to maximize the weighted sum
$\sum_{i=1}^n \lambda_i\, f_i = \tr[\rho_T\, \sum_{i=1}^n \lambda_i\, F_i]$.
Since a linear combination of Gaussian operators does in
general not have Gaussian eigenfunctions, it turns out that 
the optimal cloners with respect to single-clone fidelities
are \emph{not} Gaussian. 
For simplicity, we restrict the following discussion to the 1-to-2 cloning
problem. In this case the maximum of the weighted sum of single-copy fidelities
$\lambda_1 f_1+\lambda_2 f_2 = \tr[\rho_T\, F]$ is the largest eigenvalue of
the operator 
\begin{equation}
\label{sum} 
F = 
\lambda_1\, e^{-(Q_1^2+P_2^2)/2} + \lambda_2\, e^{-(Q_2^2+P_1^2)/2}. 
\end{equation} 
A simple numerical method to
find this eigenvalue is to iterate $\phi_{n+1}=F\phi_{n}/||F\phi_{n}||$.
Varying the weights $\lambda_i$ yields the fidelity pairs $(f_1,f_2)$
along the solid curve in Fig.~\ref{pic}. 
In comparison, the best Gaussian cloners are given by rotation
invariant Gaussian wave functions with appropriate squeezing, and the resulting
fidelity pairs are plotted in Fig.~\ref{pic} as a dotted curve.
At the intersection with the diagonal of symmetric
fidelities lie the respective optimal cloners. For the optimal
non-Gaussian cloner,
we obtain $f_1=f_2\approx 0.6826$, which is strictly higher than 
the fidelity of the optimal Gaussian cloner, namely
$f_1=f_2=2/3$ (cf.\ \cite{CIR}).

Studying cloners which are described by highly squeezed non-Gaussian states
$\rho_T$ reveals that on the curve of optimal fidelity pairs the points with
$f_1=1$ and $f_2=1$ are approached with infinite slope \cite{KWW}. It is thus
clear that the iteration for the largest eigenvalue does not become singular.
This regime is of potential interest in quantum key distribution, since nearly
perfect clones for the legitimate recipient combined with clones of non-trivial
fidelity for the eavesdropper would be the hallmark of a successful cloning
attack. On the other hand, the potential room for this regime is tiny
as it is already proven that Gaussian attacks are optimal for a large class
of quantum key distribution protocols where the channel is probed via
second-order moments of the quadratures \cite{GC04}.

\paragraph{Optical implementation.}

\begin{figure}
\includegraphics[width=0.8\columnwidth]{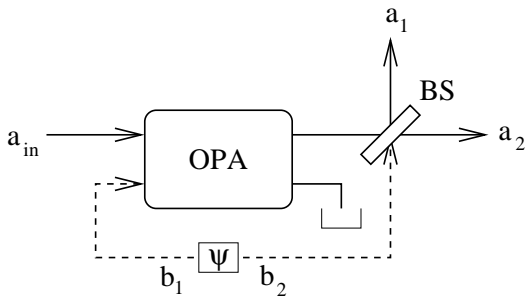}
\caption{Optical scheme of a displacement-covariant cloner.  The input mode
$a_\text{in}$ is injected on the signal mode of an optical parametric amplifier
(OPA) of gain 2, the idler mode being denoted as  $b_1$. After amplification,
the signal mode is divided at a balanced beam splitter (BS), resulting in two
clones  in modes $a_1$ and $a_2$.  The second input mode of the beam splitter
is noted $b_2$. If both $b_1$ and $b_2$ are initially in the vacuum state, the
corresponding cloner is the Gaussian cloner of \cite{L,CIR,CI}. In contrast, 
if we inject a specific two-mode state $\ket\psi$ into $b_1$ and $b_2$,  we can
generate the whole set of displacement-covariant cloners, in particular the
non-Gaussian optimal one.}
\label{implementation}
\end{figure}

The Gaussian symmetric cloner can be realized by linear amplification of the
input state, followed by distributing the output state into the two clones with
a balanced beam splitter \cite{BCILM}. This corresponds to the setup shown in
Fig.~\ref{implementation} where the idler mode of the amplifier ($b_1$) and the
second input mode of the beam splitter ($b_2$) are both initially in the vacuum
state. Let us now analyze the cloning transformation that results from
injecting an arbitrary two-mode state at modes $b_1$ and $b_2$. If the
intensity gain of the optical parametric amplifier is 2, the modes where the
two clones emerge are related to the input modes via the canonical
transformation
\begin{equation}\begin{split}
a_1 &= a_\text{in} + (b_1^\dagger + b_2)/\sqrt{2}  \\
a_2 &= a_\text{in} + (b_1^\dagger - b_2)/\sqrt{2}.
\end{split}\end{equation}
From this expression, it is straightforward to check that the
underlying cloner is displacement covariant. Moreover,
if the input is in the vacuum state $\rho=\ket0\bra0$, the single-clone
fidelities amount to expectation values of the observables
\begin{equation}\begin{split}
F_1 &= e^{-(Q_1+Q_2)^2/4-(P_1-P_2)^2/4} \\
F_2 &= e^{-(Q_1-Q_2)^2/4-(P_1+P_2)^2/4}
\end{split}\end{equation}
where $(Q_1,P_1)$ and $(Q_2,P_2)$ are the canonically conjugate field
quadratures of modes $b_1$ and $b_2$, respectively. This exactly coincides with
expression (\ref{sum}) up to a symplectic rotation, namely a beam splitter
transformation.

Consequently, the problem of finding the optimal cloner
reduces to finding the eigenstate with highest eigenvalue
of $\lambda_1 F_1 + \lambda_2 F_2$, that is, to find the optimal
bimodal state $\ket\psi$ to be injected in modes $b_1$ and $b_2$.
Note that if $\ket\psi$ is an EPR state, i.\,e.\ a suitable infinitely
squeezed state \cite{infsqueeze}, then this corresponds to the two extreme
points of the solid curve in Fig.~\ref{pic}.
The symmetric case $\lambda_1 = \lambda_2$ is obtained by choosing
\begin{equation}
\ket\psi = \sum_{n= 0}^\infty c_n \ket {2n} \ket {2n}
\end{equation}
where $\ket n$ are Fock states and the probability amplitudes $c_n$ correspond
to the dominant eigenstate of $F_1+F_2$. Truncations of this state to finite
photon numbers correspond to sub-optimal cloners: keeping only the vacuum term
$n=0$ we get the optimal Gaussian cloner with fidelities 2/3, while allowing
for $n\le 2$ yields the higher fidelities $f_1=f_2\approx 0.6801 >2/3$. The
experimental realization of this cloner does not seem unrealistic, given the
recently proposed schemes for conditionally preparing arbitrary bimodal states
of light based on linear optics \cite{KLD02}. In the limit $n\to \infty$, we
arrive at the optimal cloner with $f_1=f_2\approx 0.6826$.

Independent studies of the cloning fidelities of coherent states in
finite-dimensional Hilbert spaces and their numerical extrapolation has
indicated that the optimal fidelity ranges between 2/3 and 0.699, which is in
accordance with our result \cite{DKW}.

\paragraph{Optimal classical cloning.}
Let us finally consider a classical 1-to-$\infty$ cloning map $T$ which is
realized by measuring and repreparing the system. From the line of
arguments above, $T$ can be assumed to be covariant. Since
composing this cloner with time reversal $\tau$ leads to a
completely positive map, $\tau\circ T_{\ast}(W_{p,q}) =
\chi_T(\sqrt{2}\,p, \sqrt{2}\,q)\, W_{-p,q}$ with $\chi_T(p,q)$
the characteristic function of a state. Computing the fidelity for
coherent input states immediately yields:
\be
    f_\text{classical}(T,|0\rangle\langle0|) = \frac{1}{2}\, \tr\bigl[\rho_T\,
        |0\rangle\langle0|\bigr] \le \frac{1}{2}.
\ee
The bound is reached by a heterodyne measurement and repreparation of coherent
states, i.\,e.\ by a Gaussian scheme. This limit can not even be surpassed with
the assistance of PPT bound entanglement \cite{PPT}, since the respective maps
are included in the above argumentation.
In the case of an unassisted measure-and-prepare scheme an independent proof
was recently given in \cite{HWC04}.

\vspace{0.2ex}
\begin{acknowledgments}
We thank J.\,I.~Cirac, J.~Eisert, J.~Fiurasek, S.~Iblisdir,
and D.~Schlingemann for interesting discussions.

We acknowledge EU funding under project COVAQIAL (FP6-511004).
NJC and PN acknowledge financial support from the Communaut\'e Fran\c caise
de Belgique under grant ARC 00/05-251, from the IUAP programme of the 
Belgian government under grant V-18.
\end{acknowledgments}

\end{document}